\documentclass[prb,preprint,showpacs,superscriptaddress]{revtex4}
\usepackage{graphicx}

\begin{document}
\title{\bf Computational and experimental imaging
of Mn defects\\ on GaAs  (110) cross-sectional surface}
\author{A. Stroppa}
\email{astroppa@ts.infn.it}
\altaffiliation[Presently at: ]{Institute of Material Physics,
University of Vienna, Sensengasse 8/12, A-1090 Wien, Austria and
Center for Computational Materials Science (CMS), Wien, Austria}
\affiliation{Dipartimento di Fisica Teorica, Universit\`a di
Trieste,\\ Strada Costiera 11, 34014 Trieste, Italy}
\affiliation{CNR-INFM DEMOCRITOS National Simulation Center, via Beirut 2-4, 
34014 Trieste, Italy}
\author{X. Duan}
\altaffiliation[Presently at ]{School of Physics,
The University of Sydney, NSW 2006 Australia}
\affiliation{Dipartimento di Fisica Teorica, Universit\`a di
Trieste,\\ Strada Costiera 11, 34014 Trieste, Italy}
\affiliation{CNR-INFM DEMOCRITOS National Simulation Center, via Beirut 2-4, 
34014 Trieste, Italy}
\author{M. Peressi}
\email{peressi@ts.infn.it}
\affiliation{Dipartimento di Fisica Teorica, Universit\`a di
Trieste,\\ Strada Costiera 11, 34014 Trieste, Italy}
\affiliation{CNR-INFM DEMOCRITOS National Simulation Center, via Beirut 2-4, 
34014 Trieste, Italy}
\author{D. Furlanetto}
\affiliation{CNR-INFM TASC National Laboratory, Area Science Park, 34012
Trieste, Italy}
\affiliation{Dipartimento di Fisica and
Center of Excellence for Nanostructured Materials, CENMAT,\\
Universit\`a di Trieste,
 via A. Valerio 2, 34127 Trieste, Italy}
\author{S. Modesti}
\affiliation{CNR-INFM TASC National Laboratory, Area Science Park, 34012
Trieste, Italy}
\affiliation{Dipartimento di Fisica and
Center of Excellence for Nanostructured Materials, CENMAT,\\
Universit\`a di Trieste,
 via A. Valerio 2, 34127 Trieste, Italy}

\date{\today}

\begin{abstract}
We present a combined experimental and computational study of the (110)
cross-sectional surface of Mn $\delta$-doped GaAs samples.
We focus our study on three different selected Mn defect configurations
not previously studied in details, namely surface interstitial Mn, 
isolated and in pairs, and  substitutional 
Mn atoms on cationic sites (Mn$_{\rm Ga}$) in the first subsurface layer.
The sensitivity of the STM images to the specific local environment allows to distinguish between
Mn interstitials with nearest neighbor As atoms (Int$_{\rm As}$) rather than Ga atoms (Int$_{\rm Ga}$),
and to identify the fingerprint of peculiar satellite features around subsurface substitutional
Mn.
The simulated STM maps for Int$_{\rm As}$, both isolated and in pairs, and
Mn$_{\rm Ga}$ in the first subsurface layer
are consistent with some experimental images hitherto not fully characterized.
\end{abstract}

.

\maketitle

\newpage
\section{Introduction}
Mn-doped GaAs\cite{Takamura,strong,reviewOhno,Jungwirth}  has
attracted considerable attention among the
diluted magnetic semiconductors for its possible application in the emerging
field of \emph{spintronic}.\cite{spintronic1,spintronic2,spintronic3} 
Although other materials such as
ferromagnetic metals and alloys, Heusler alloys, or magnetic oxides
seem to be promising candidates for spintronic devices, 
the diluted magnetic semiconductors and Mn-doped GaAs in particular
are of tremendous interest in that
they combine magnetic and semiconducting properties and allow
an easy integration with the well established semiconductor technology.
Besides possible spintronic applications, characterizing and understanding 
the properties of Mn defects in GaAs is a basic research problem which is
still debated.

The growth conditions and techniques affect
the solubility of Mn in GaAs, which is in general rather limited,
and its particular defect configurations, thus determining 
the magnetic properties of the samples.\cite{interst5,annea1,annea2,PRLMN,defectth}
The highest Curie temperature T$_{c}$ reachable
for Mn-doped GaAs  up to  few years ago was 110
K,\cite{accept2}  rather low for practical technological
purposes. Intense efforts
have been pursued in the last years in order to understand the
physics of this material and to improve its quality and efficiency.
Out-equilibrium growth techniques\cite{spintronic1,Takamura} 
have enabled to increase
the solubility of Mn and   the Curie temperature; 
post-growth annealing of epitaxial samples  at temperatures only slightly above
the growth temperature has been particularly successfull.\cite{annea1,annea2,Jungwirth2005}
Nowadays, $\delta$-doping is used as an alternative  to the
growth of bulk Mn$_{x}$Ga$_{1-x}$As,\cite{Delta,delta1} allowing to obtain
locally high dopant concentrations  and, remarkably, an
important enhancement of T$_{c}$, up to about  250 K.\cite{delta2,HighTC}

For further improvements
it is essential to investigate the different configurations of Mn impurities
and their effect on the magnetic properties of the system.
The most common and widely studied Mn configuration is substitutional in the
cation sites (Mn$_{Ga}$), with Mn acting  as a 
hole-producing acceptor.\cite{delta2} To a less extent, Mn can
also occupy interstitial sites, in particular tetrahedral ones. 
In such a case, it is expected to strongly modify the magnetic
properties, acting as an electron-producing donor and hence destroying
the free  holes and hindering ferromagnetism.\cite{ZungerPRB2003}

Interstitials have not been fully characterized
up to now, although their existence has been suggested in different 
situations.\cite{interst5,vanGisbergen,annea1,annea2,GlasPRL2004,MahieuAPL2003,EdmondsPRB2005,Jungwirth2005,Wu,PRLMN,ErwinPRL2002,Holy2006}
For instance, the enhancement of the Curie temperature after
post-growth annealing  has been attributed to the reduction of interstitial
defects with their out diffusion towards the surface.\cite{PRLMN}
It has been suggested that
interstitial sites are highly mobile and
could be immobilized when  adjacent
to substitutional Mn$_{Ga}$, thus forming compensated pairs
with antiferromagnetic coupling.\cite{Blinowski2003}
A first identification of interstitial Mn 
dates back to almost fifteen years ago
by electron paramagnetic resonance (EPR).\cite{vanGisbergen} 
Very recently EPR
spectra from variously doped and grown samples of Mn-doped epitaxial
GaAs have allowed to identify the presence
of ionized Mn interstitials at concentrations as low as 0.5\%, although not
providing  details about the specific local environment of the interstitial
site.\cite{Weiers2006}
Recent X-ray absorption near edge structure (XANES) and
extended x-ray absorption fine structure (EXAFS) spectra
in Mn $\delta$-doped GaAs samples
suggest that Mn occupy not only  substitutional Ga sites but
\emph{also} interstitial sites, mainly in case of Be co-doping.\cite{DACAPITO}

Cross-sectional Scanning Tunneling Microscopy (XSTM) allows a direct
imaging of the electronic states and can be used to characterize
the impurities near the cleavage surface.\cite{Feenstra}
In recent years several XSTM studies of Mn-doped GaAs samples
have been performed but without a complete consensus on the
defects characterization.\cite{MahieuAPL2003,Mikkelsen,review-Mikkelsen2005,Sullivan,Yakunin1,Yakunin2,SingleMnImpurity,KitchenNAT06,GleasonAPL2005}
We stress that most of XSTM studies mainly concern Mn$_{x}$Ga$_{1-x}$As
alloys and have identified mainly substitutional Mn defects.
$\delta$-doped samples have been investigated by Yakunin et al.,\cite{Yakunin2}
who pointed out the advantage that in such samples it is easy to discriminate
Mn related defects from other defects.

From the theoretical point of view, numerical works
have been also focused  mainly on the simulation of
 XSTM images of substitutional impurities on uppermost surface
layers.\cite{Mikkelsen,Sullivan,Yakunin1,Yakunin2}
A complete and detailed investigation of
interstitial impurities as they can appear on the exposed cleaved surface
is still lacking
thus preventing the possibility of a comprehensive interpretation of
all the available experimental XSTM images.

Mn $\delta$-doped (001) GaAs samples recently grown at TASC Laboratory in
Trieste and analyzed with XSTM on the (110) cleavage surface have shown 
several Mn related features (see Fig. 1). Some of them have already be studied 
by other groups, like the asymmetric cross-like (or butterfly-like) structures 
marked by A in Fig.~1(a), attributed to Mn acceptors a few atomic layers below 
the surface.\cite{Yakunin1}
Some other features, such those marked by B, or those of Fig 1(b), 
have not been yet assigned to specific Mn configurations.  In 
order to identify the kind of Mn defects that cause them we have performed new 
density functional simulation of cross-sectional XSTM images focusing on three 
selected defect configurations not yet fully studied, but whose presence 
cannot be excluded in real samples. In particular, we focus our attention on 
interstitial surface configurations, both individual as well as in pairs. We 
have also considered Mn$_{Ga}$ 
on the first layer below the surface and compared 
all the simulations with the experimental maps.

\section{Experimental details}
Mn  $\delta$-doped samples were grown by molecular beam epitaxy on GaAs(001)
in a facility which includes a growth chamber for III-V materials and a
metallization chamber. After the growth of a Be doped buffer at 590$^o$C
and of an undoped GaAs layer 50 nm thick at 450$^o$C with an As/Ga beam
pressure ratio of 15, 
the samples were transferred in the
metallization chamber where a submonolayer-thick Mn layer was
deposited at room temperature at the rate of 0.003 monolayer/s. An
undoped GaAs cap layer was subsequently grown at 450$^o$C.  
This procedure was repeated in order to have three $\delta$-doped Mn 
layer of 0.01, 0.05 and 0.2 monolayers in the same sample. 
During the transfers and the Mn deposition the vacuum was always 
better than $2\times 10^{-8}$ Pa. 
The 0.1 mm thick wafers
containing the Mn layers were cleaved in situ in a ultra high vacuum
STM system immediately prior to image
acquisition to yield atomically flat, electronically unpinned \{110\}
surfaces containing the [001] growth direction and the cross section
of the $\delta$-doped layer. The XSTM image presented in 
Fig.~\ref{fig:exp} and the others shown in this paper have been
acquired from a  $\delta$-doped Mn layer of 0.2 monolayers with W tips.

The densities of the features observed by XSTM near each Mn layer
were approximately proportional to the Mn coverage of the $\delta$-doped
layer in the range 0.01-0.2 monolayer. No trace of contaminants was
observed by in situ x-ray photoemission spectroscopy  after the
transfer in the metallization, after the Mn deposition, and after the
transfer in the growth chamber. For these two reasons we attribute the
features observed by XSTM to the Mn atoms, and not to defects or
contaminants caused by the growth interruption and  transfers
between the chambers. The density of the defects caused by these steps
should not depend on the Mn coverage, contrary to what we observe. 
Moreover, a sample was grown with the same procedure described above,
including the transfers between the chambers, but without the Mn
deposition.  The photoluminescence spectra of this sample are
undistinguishable from that of a good undoped GaAs epitaxial layer
grown without transfers between the chambers. This confirms that the
transfers do not introduce an appreciable amount of defects.

\section{Theoretical approach}
Our numerical approach is based on spin-resolved Density Functional
Theory (DFT) using  the
{\em ab-initio} pseudopotential plane-wave method {\tt PWscf} code of
the {\em Quantum ESPRESSO} distribution.\cite{pwscf} 
Cross-sectional surfaces are studied using supercells with slab geometries,
according to a scheme previously used,\cite{XM} with 5 
atomic layers and a vacuum region equivalent to 8 atomic layers.
Mn dopants are on one surface, whereas the other is passivated with hydrogen.
For a single Mn impurity we use a  4$\times4$ in-plane periodicity
corresponding to distances between the
Mn atom and its periodic images of 15.7 \AA\
along the [1$\bar{1}$0] and 22.2 \AA\ along [001].
No substantial changes in the XSTM images 
have been found using a  6$\times$4 periodicity, which
has been instead routinely used when considering interstitial complexes.
Other details on technicalities can be found in 
Ref.~\onlinecite{StroppaMatSciEngB}.

In our study, 
we have mainly focused on the  Local Spin Density Approximation (LSDA)
 for the exchange-correlation  functional. 
 An ultrasoft pseudopotential is used for Mn atom, considering
semicore $3p$ and $3s$ states kept in the valence shell while
norm-conserving pseudopotentials have been considered for Ga and As atoms. The
3d-Ga electrons are considered as part of core states.\cite{pseudi,Debernardi}

Tests beyond LSDA (with Generalized Gradient Correction
and LSDA+U methods)  have not shown any
 substantial difference in the features of the XSTM maps.
As a further check, we have
also simulated ionized substitutional  
Mn$_{Ga}$ (with charge state equal to $1-$) on surface 
and in the first subsurface layer as well as ionized interstitial Mn (with
charge state equal to $2+$) on surface layer.
Neither the former nor the latter 
simulated XSTM maps show significant differences with respect to the neutral 
cases. We address the reader to a future pubblication for details.\cite{future}

 The XSTM images are simulated using the model of
 Tersoff-Hamann,\cite{Ters1,Ters2} where the tunneling current is proportional 
to the Local Density of
States (LDOS) at the position of the tip, integrated
in the energy  range between the Fermi energy $E_{f}$ and $E_{f}+eV_b$, 
where $V_b$
 is the  bias applied to the sample with respect to the  tip.
The position of the Fermi
 level is relevant for the XSTM images. In general,  $E_{f}$
strongly depends on the concentration of dopants: this is contrivedly
large in our simulations even in the case of a single Mn dopant per
supercell. Therefore to overcome this problem we fix  $E_{f}$ according to
the experimental indications: in order to account for the $p$-doping in the real
samples, we set $E_{f}$
 close to the Valence Band Maximum (VBM).
The VBM in the DOS of the Mn-doped GaAs can be exactly identified 
by aligning  the DOS projected onto surface  atoms
 far from the impurity with the one of the clean surface.
In any case, the comparison between experiments and simulations must be taken  with some
caution, due to the possible differences 
in the details entering in the determination of the XSTM image, such as 
tip-surface separation,  precise value of the bias voltage and 
position of $E_f$, surface band gap.\cite{note}  

\section{Surface Mn interstitials}

We first focus on interstitial dopant configurations,  
Int$_{As}$ and Int$_{Ga}$.
Throughout this work we have considered only \emph{tetrahedral}
interstitial position, since it is known from bulk calculations that
the total energy corresponding to the \emph{hexagonal} interstitial
site is higher by more than 0.5 eV.\cite{Maka2,PRLMN,condmat}  The tetrahedral
interstitial site in the ideal geometry has four nearest-neighbor
(NN) atoms at a distance equal to the ideal host bond length $d_{1}$
and six next-nearest-neighbor (NNN) atoms at the distance
$d_{2}=\frac{2}{\sqrt 3}d_{1}$, which are As(Ga) atoms for
Int$_{Ga(As)}$, respectively. At the ideal truncated (110) surface,
the numbers of  NNs and NNNs reduce to three (2 surface atoms and 1
subsurface atom) and four (2 surface atoms and 2 subsurface atoms)
instead of four and six respectively.

In the uppermost panels of Fig.~\ref{fig:int} we show a ball and stick side and
top  view of the relaxed Int$_{As}$ and Int$_{Ga}$ configurations.
 In the relaxed structure, due to symmetry breaking because of the surface and
 the consequent buckling of the outermost surface layers, the NN and  NNN bond lengths are
no longer equal. Furthermore, some relaxed NNs bond lengths turn
out to be longer than NNNs ones.
In the following, we do not  distinguish among NN and NNN
atoms: they are simply referred as neighbor \emph{surface} or
\emph{subsurface} atoms, as shown in the Figure.

The two relaxed configurations slightly differ in energy,  by $\sim$
130 meV/Mn atom, in favour of Int$_{Ga}$. 
This is at variance with the bulk case studied in  the literature,
where it has been found that  Int$_{As}$ is favoured:
for neutral state, the energy difference is actually so small 
(5 meV/Mn atom)\cite{Maka2}  that it is not meaningful,
but it goes up to 350 meV 
in case of interstitial Mn with 2+ charge state.\cite{PRLMN}

After optimization of the atomic positions, 
 sizeable displacements from the ideal zinc
blende positions occur for the Mn impurities and their 
surface and subsurface neighbors; small relaxations effects are still present
in the third layer, in both configurations.
In Int$_{As}$, with respect to the ideal (110) surface plane,
Mn relaxes outward by $\sim$ 0.06 \AA\ and
As$_{surf}$ (As$_{subsurf}$) move upwards (downwards). On the other
hand, the Ga atoms (both on surface and subsurface) are shifted
towards the bulk. In Int$_{Ga}$, Mn relaxes inward by $\sim$ 0.32
\AA; the Ga$_{surf}$ and Ga$_{subsurf}$ atoms are displaced
downwards and the As$_{surf}$ (As$_{subsurf}$) atom moves upwards
(downwards). The interatomic distances between Mn and the nearest
atoms  are in general longer by more than 2-3 \%
than ideal values
(details in Ref.~\onlinecite{StroppaMatSciEngB}).

The simulated XSTM images
of Int$_{As}$ (left) and Int$_{Ga}$  (right)  configurations
at negative and positive bias voltages (from $-$ 2.0 V to $+$2.0 V)
are shown in the lower panels of Fig.~\ref{fig:int}.
In Int$_{As}$,  Mn appears as an
additional bright spot at negative bias voltage ($V_b$=$-$1 V), 
slightly elongated in the [001] direction 
and located near the center of the surface unit cell identified by
surface As atoms. The As$_{surf}$ atoms close to Mn appear less bright than
the others.  These features are similar changing $V_b$ from $-$1 to $-$2 V.

In the empty states image at $V_b$=1 V Mn appears again as an elongated bright
spot. The
underlying cation lattice is only barely visible at this bias
voltage. The very bright XSTM feature originates from
the Mn $d$ minority states and a strong peak of
Ga$_{surf}$ majority states.\cite{StroppaNuovoCim}
At $V_b$=2 V, this feature is still well visible,
as well as another
region brighter than the underlying cationic sublattice
in correspondence of  $As_{surf}$ atoms  neighbor to Mn,
suggesting a contribution coming from the
hybridization between  Mn-$d$ and As$_{surf}$-$p$  states.

In Int$_{Ga}$ configuration, at negative voltage, Mn appears as an almost
circular bright spot located in between two surface As atoms
 adjacent along the [001] direction.  
At positive bias voltages, the two Ga$_{surf}$ atoms neighbor to Mn
 appear very bright with features extending towards Mn in a ``v''-shaped form
and the atoms in the neighborhood also look brighter than normal. These
features remain visible by increasing the positive bias voltage up to 2 eV.
Remarkably the empty states images of Mn are quite different for the
two interstitial configurations, making them clearly distinguishable by XSTM analysis. 
 Some features in the experimental XSTM images appear as bright spots both at 
positive and negative bias voltages. These spots lie along the  [001] Ga rows  
and between the [1-10]  Ga columns  at positive bias voltage 
 (see Fig.~\ref{fig:exp}(b)). 
 Their location with respect to the surface Ga lattice and the comparison with
 the simulated images allow to identify them as Int$_{As}$ Mn atoms.

The numerical simulation gives easily  informations on the
magnetic properties of the system.
The total and absolute magnetization,
calculated from the spatial integration of
the difference and the absolute difference respectively 
between the majority and minority electronic charge distribution,
are different in the 
two configurations: 4.23 and 4.84 $\mu_B$ for Int$_{As}$ and 3.41 and 4.71
 $\mu_B$ for Int$_{Ga}$ respectively. 
These differences indicate in both cases
the presence of region of negative spin-density and 
a clear dependence  of the induced magnetization on the local Mn environment.
The individual atomic magnetic moments can be calculated
as the difference between the majority and minority atomic-projected charges.
In Int$_{As}$, Mn magnetic moment is 3.96 $\mu_B$, almost integer,
corresponding to the presence of a gap in the Mn-projected minority
density of states.
Mn magnetization is slightly lower in Int$_{Ga}$ (3.67 $\mu_B$).
In both cases they are significantly larger compared to the bulk case,
indicating a surface induced enhancement.
The analysis of spin-polarization induced by interstitial Mn on its
nearest neighbors shows in both cases an antiferromagnetic Mn--Ga coupling and 
 a smaller ferromagnetic Mn--As coupling: more precisely,
the magnetic moments induced on surface Ga atoms neighbors to Mn are
equal to $-$0.14 and $-$0.17 $\mu_B$ in Int$_{As}$ and Int$_{Ga}$ respectively,
whereas those induced on surface or subsurface As atoms neighbors to Mn
are positive and at most equal to 0.05  $\mu_B$.
We address the reader to Ref. \cite{StroppaMatSciEngB} for further details.

In the experimental images of Mn $\delta$-doped GaAs samples
we often observe two spots close one each other at a distance of about 8 \AA,
as reported  in  Fig.~\ref{fig:doublet-exp-th-2} (larger panel).
The simulated image of two  Int$_{As}$ atoms 
separated by a clean surface unit cell along (1$\overline{1}$0),
partially superimposed, reproduces the main features of this experimental 
image,
and it is basically a superposition of  images  of individual Int$_{As}$
(elongated bright spot each one,  with major axis along the [001] direction,
and a surrounding darker region).

\section{
  Substitutional Mn defects in the first subsurface layer} 

Another typical feature present in the experimental XSTM maps is a 
bright spot visible at positive bias voltages with two satellite features
forming a triangular structure, as shown 
in Fig. 1(a) (feature B) and in Fig. 4 in the lower panels. This
feature seems similar to that caused by the arsenic antisite defect
(As on Ga) in GaAs.\cite{FeenstraPRL1993,Mathieu} However in the arsenic
antisite defects the satellites are visible only at negative sample
bias, while the defect that we observe in the Mn layers shows
satellite only in the positive bias images. On the other hand there is
a clear resemblance of the defect B (Fig. 1(a) and Fig. 4) with the
 simulated image of a  substitutional Mn$_{\rm Ga}$ atom
in the first subsurface layer
shown in the  panels partially superimposed to the experimental images.
It can be seen at $V_b<0$ 
a deformation of the surface As rows in correspondence of 
the Mn impurity below, and, even more remarkably, the peculiar satellite bright
features on two neighboring surface As stoms at $V_b>0$ giving rise to a 
triangular-shaped image.
Therefore we attribute
the defect B to substitutional Mn Ga atoms in the first subsurface
layer.

Finally, we discuss our findings in comparison with some relevant results 
present in the literature.
The comparison of our simulations with those of 
 Sullivan et al.\cite{Sullivan} is possible only for the 
isolated Mn$_{\rm Ga}$ in the first subsurface layer 
at negative bias voltage:
in such a case the simulated images show similar features.
The corresponding image at positive bias is not reported and
other configurations are not comparable. 

The XSTM imaging of  substitutional Mn
is reported with more details  by Mikkelsen 
et al.,\cite{Mikkelsen,review-Mikkelsen2005}
where both the simulated maps for surface and subsurface Mn$_{\rm Ga}$
 and the experimental ones attributed to
this impurity configuration are shown at negative and positive bias,
thus allowing for a more complete comparison. 
The images for Mn$_{\rm Ga}$ in the first subsurface layer have  a good
resemblance with ours, a part from the 
satellite features that we have identified at positive bias 
on neighbor As atoms which are not 
present in their images, neither in the simulated nor in the experimental one.
More precisely, we notice that their simulated surface area is too
small to make such satellite features visible.
The simulated images for surface Mn$_{\rm Ga}$ are also 
similar to ours and, like ours, 
not corresponding to any experimental feature.\cite{future}
This leads to the conclusion that
the presence of substitutional Mn in the first layer of the exposed surface
is very unlikely.

Mikkelsen et al.  reported also the simulation of surface interstitial Mn  in 
their Fig. 3(d),\cite{review-Mikkelsen2005} that according to our understanding
on the basis of the symmetry planes should correspond to
 Int$_{\rm Ga}$,  although not  explicitely indicated. 
Their images are similar to ours for the same configuration. They rule out the 
presence of interstitials since these images are not compatible with
experiments, at variance with our findings concerning Int$_{\rm As}$.
It should be noted however
that we observe the Int$_{\rm As}$ features in the experimental samples
 only in the first few 
hours after the sample cleavage. They disappear for longer times, probably 
because of  surface contamination or diffusion. 

 Kitchen et al.\cite{SingleMnImpurity,KitchenNAT06}  
report experimental images for Mn adatoms at the GaAs (110) surface
with highly  anisotropic  extended star-like feature, attributed to a single
surface Mn acceptor.
Interestingly, these images are compatible with our simulated
surface  Mn$_{\rm Ga}$, not show here.\cite{future}
A resemblance with 
our empty state image for Int$_{As}$ (see Fig.~\ref{fig:int} at $V_b$=+2 V)
is instead only apparent because the  mirror symmetry plane is different. 

An anisotropic, crosslike feature in XSTM image is reported also Yakunin
et al.\cite{Yakunin1} and, from comparison with an
envelope-function, effective mass model and  a tight-binding model, it is
attributed to  a hole bound to an individual Mn acceptor
lying well below the surface. We observe similar 
feature of different sizes (see Fig. 1), the smallest of them are those 
reported in Fig. 4, that we identify as Mn$_{Ga}$ 
in the first subsurface layer. 

A part from different details, our simulated images for  surface and
subsurface Mn$_{\rm Ga}$ are compatible with such crosslike features,
although
experimental and simulated images reported therein concern substitutional 
 impurities located more deeply subsurface than those we have considered. 
Crosslike features are observed  even at very short Mn-Mn spatial
separations.\cite{Yakunin2}

\section{Conclusions}
We have reported  a combined experimental and first-principles
numerical study of XSTM images of the (110)
cross-sectional surfaces of Mn $\delta$-doped GaAs samples.
We suggest an identification of three typical configurations
observed in the experimental sample on the basis of a comparison of numerical
prediction and observed images both at negative and positive applied bias.
(i) Some structures observed can be identified 
as surface Mn interstitial with As nearest neighbors, 
on the basis of their position with respect to the surface lattice and the 
comparison with the simulated images.
At variance, there is no evidence in the experimental samples of 
Mn interstitial
with Ga nearest neighbors, whose  XSTM imaging according to our numerical
simulations would correspond to very different features.
(ii) Besides isolated configurations, also pairs of Mn interstitials with
As  nearest neighbors are clearly observed and identified.
(iii)
Subsurface substitutional Mn$_{\rm Ga}$ atoms in the first subsurface layer
can also be unambigously
identified in the experimental images by a main bright spot corresponding to 
the dopant and from  peculiar satellite features on two neighboring As atoms
which are clearly observed in the
experimental images and predicted by simulations.

\section{Acknowledgments}
Computational resources have
been partly  obtained within the ``Iniziativa Trasversale di Calcolo
Parallelo'' of the Italian {\em CNR-Istituto Nazionale per la Fisica
della Materia} (CNR-INFM) and partly within the agreement between the
University of Trieste and the Consorzio Interuniversitario CINECA
(Italy). We thank A. Franciosi, 
S. Rubini and coworkers for the preparation of the sample
and fruitful comments and discussions; 
A. Debernardi for his help in the pseudopotential generation and
for useful discussions.
Ball and stick models and simulated images are obtained with the package
XCrySDen.\cite{XCrySDen}

\clearpage

\begin{center}
\begin{figure}[!hbp]
\caption{
(a) Experimental (110) XSTM image of a 0.2 monolayer Mn $\delta$-doped 
layer in GaAs at a sample bias voltage of  1.7 eV. This image has not been 
corrected for the drift of the sample. (b)  XSTM image of a Mn 
related structure at the bias voltage of $-$1.4 eV (left) and $+$1.9 eV 
(right).  The white lines show the [001] Ga atomic rows.}
\label{fig:exp}
\end{figure}
\end{center}

\begin{center}
\begin{figure}[!hbp]
\caption{
Isolated  Mn interstitial dopants on
GaAs(110) surface, with As nearest neighbors (Int$_{As}$, left) and
Ga nearest neighbors (Int$_{Ga}$, right). Upper panels:
ball-and-stick model of the relaxed surface, top and side view.
Only the three topmost layers are shown in the side view.
Black spheres are Mn, white spheres are As, grey spheres are Ga.
Lower panels: simulated
XSTM images at occupied states and empty states respectively, for
different bias voltages.
}\label{fig:int}
\end{figure}
\end{center}

\begin{center}
\begin{figure}[!hbp]
\caption{Smaller superimposed panel:
simulated XSTM image of a pair of Int$_{As}$ on GaAs(110)
surface with a relative distance of $\sim$ 8 \AA \ along the [1\={1}0]
direction at a bias voltage $V_b$=$-$2 V.
The larger panel shows an experimental image compatible with the simulation.} 
\label{fig:doublet-exp-th-2}
\end{figure}
\end{center}

\begin{center}
\begin{figure}[!hbp]
\caption{Upper smaller superimposed panels
simulated XSTM image of a  subsurface Mn$_{Ga}$  on GaAs(110)
at negative (left) and  positive (right) bias voltages.
The lower panels show corresponding  experimental 
images of the structure B (see Fig. 1(a))  
taken at sample bias voltages of $-$1.4 V (left) and 
$+$1.8 V (right) that are compatible with the simulations,
performed with voltages of  $-$1 V and +1 V.}
\label{fig:subMnGa-exp-th}
\end{figure}
\end{center}

\clearpage
\begin{center}
\begin{figure}[!hbp]
\includegraphics[scale=.65,angle=0]{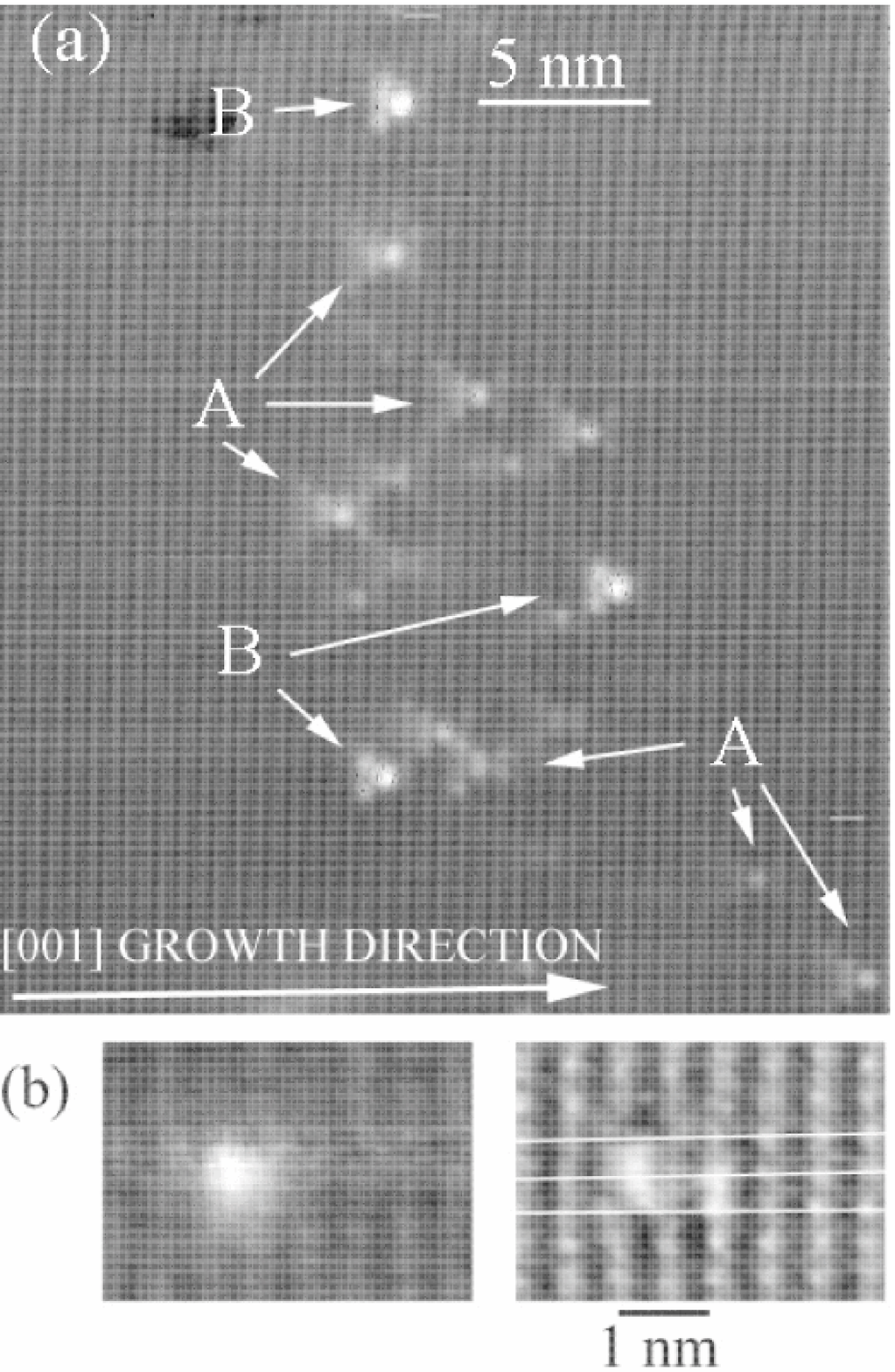}
\\Fig.~\ref{fig:exp}
\end{figure}
\end{center}

\clearpage
\begin{center}
\begin{figure}[!hbp]
\includegraphics[scale=.65,angle=0]{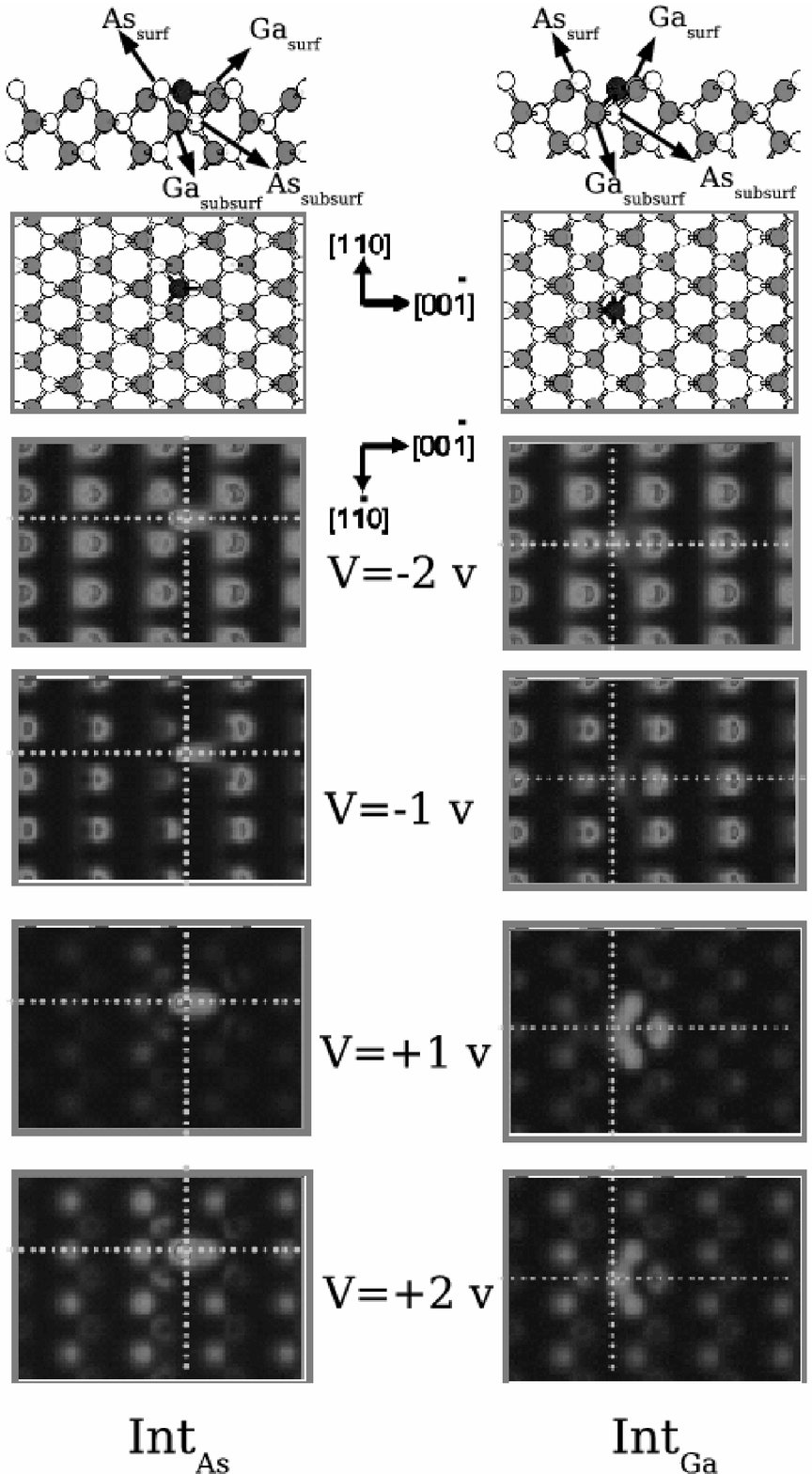}
\\Fig.~\ref{fig:int}
\end{figure}
\end{center}

\clearpage
\begin{center}
\begin{figure}[!hbp]
\includegraphics[scale=.9,angle=0]{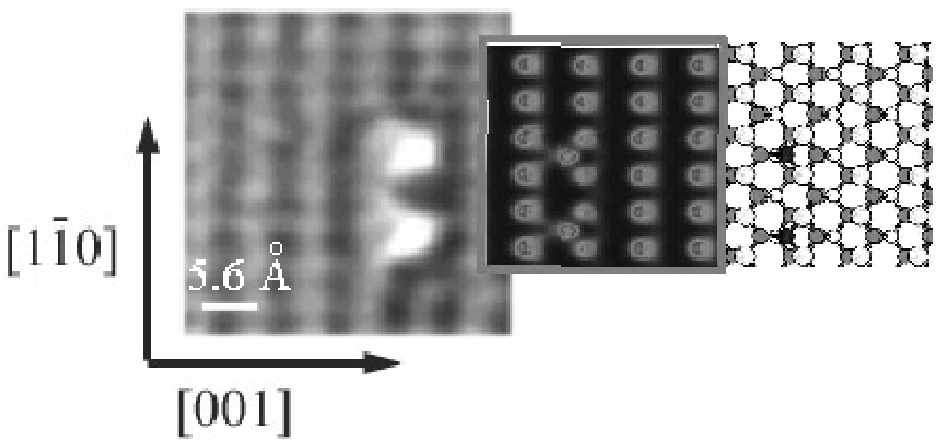}
\\Fig.~\ref{fig:doublet-exp-th-2}
\label{doublets}
\end{figure}
\end{center}

\clearpage
\begin{center}
\begin{figure}[!hbp]
\includegraphics[scale=.8,angle=0]{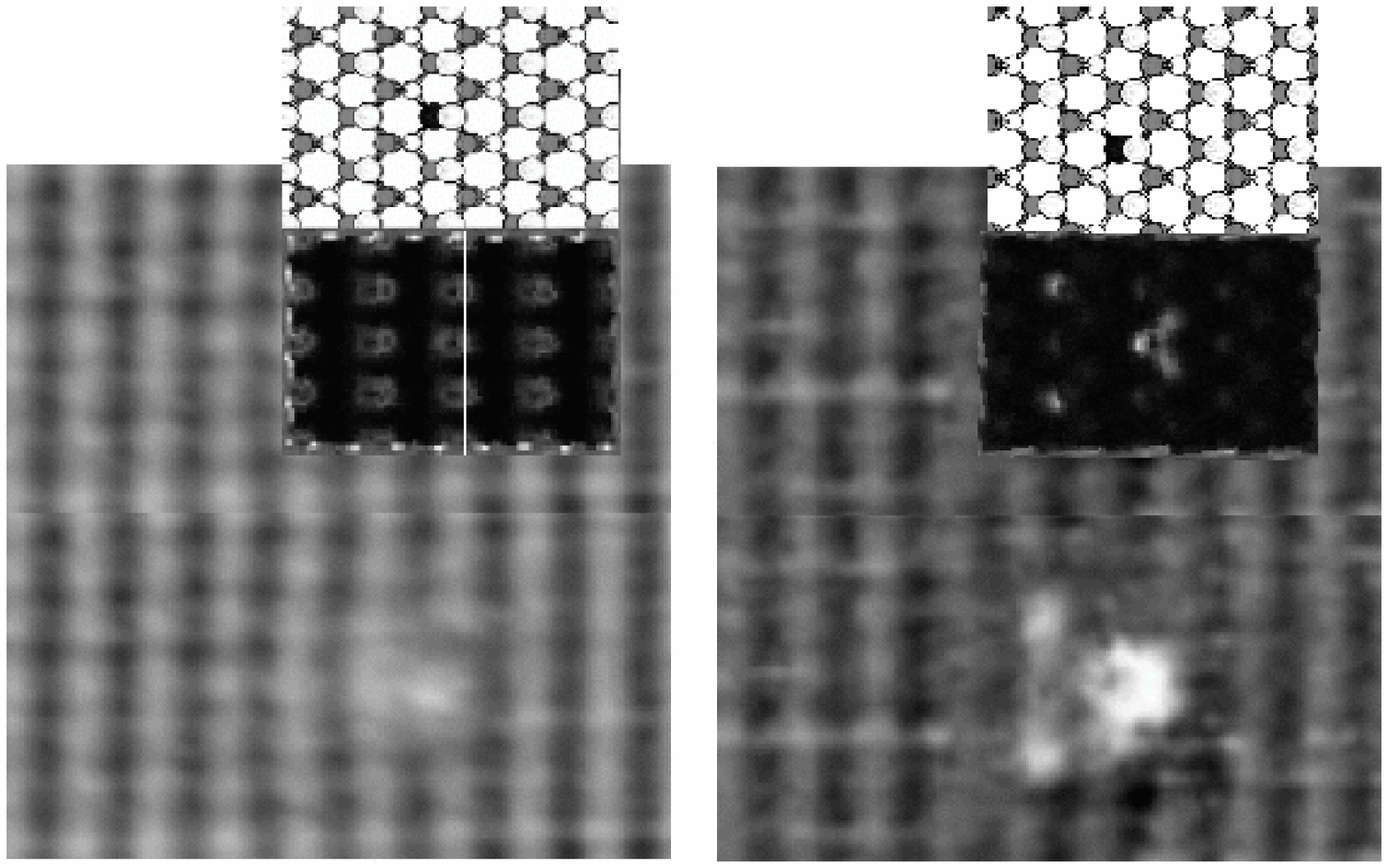}
\\Fig.~\ref{fig:subMnGa-exp-th}
\end{figure}
\end{center}



\begin{thebibliography}{99}
\bibitem{note} {However, we have checked that a variation of 
$\sim \ \pm$ 0.3 eV in the bias voltage considered in the simulated images
 does not affect their basic features.}
\end{thebibliography}
\end{document}